\documentclass[prb,aps,twocolumn,groupedaddress]{revtex4}
\usepackage{epsfig,latexsym,amssymb,amsmath,amsbsy,graphics,graphicx}

\def \beq{\begin{equation}}
\def \eeq{\end{equation}}
\def \beqarr{\begin{eqnarray}}
\def \eeqarr{\end{eqnarray}}

\begin{document}

\title{The enigma of the $\nu=0$ quantum Hall effect in graphene}

\author{S. Das Sarma}

\affiliation{Condensed Matter Theory Center, Department of Physics,
University of Maryland, College Park, MD 20742, USA}

\author{Kun Yang}

\affiliation{National High Magnetic Field Laboratory and Department
of Physics, Florida State University, Tallahassee, FL 32306, USA}

\date{\today}

\begin{abstract}

We apply Laughlin's gauge argument to analyze the $\nu=0$ quantum Hall effect observed in graphene when the Fermi energy lies near the Dirac point, and conclude that this necessarily leads to divergent bulk longitudinal resistivity in the zero temperature thermodynamic limit. We further predict that in a Corbino geometry measurement, where edge transport and other mesoscopic effects are unimportant, one should find the longitudinal conductivity vanishing in all graphene samples which have an underlying $\nu=0$ quantized Hall effect. We argue that this $\nu=0$ graphene quantum Hall state is qualitatively similar to the high field insulating phase (also known as the Hall insulator) in the lowest Landau level of ordinary semiconductor two-dimensional electron systems.  We establish the necessity of having a high magnetic field and high mobility samples for the observation of the divergent resistivity as arising from the existence of disorder-induced  density inhomogeneity at the graphene Dirac point.

\end{abstract}

\pacs{72.10.-d, 73.21.-b,73.50.Fq}

\maketitle


A single two-dimensional (2D) layer of carbon atoms forming a honeycomb lattice, i.e., a graphene layer, has unusual physical properties attracting a great deal of current interest.\cite{graphene_exp} Among its intriguing properties, the effective low-energy dispersion is linear in 2D momentum: $E=\hbar v |{\bf k}|$, where $v$, the graphene Fermi velocity, is a constant ($v\sim c/300$ where $c$ is speed of light), and electron wave functions formally obey a Dirac-like continuum equation with zero Dirac mass, rather than the Schrodinger's equation. Associated with this Dirac nature is the fact that the single-particle spectrum has a two-fold pseudo-spin (or valley) degeneracy, in addition to the usual (double) spin degeneracy. Thus the low-energy spectrum of graphene is made of particle or hole excitations near a double-cone Fermi surface; the apex of the cones are the Dirac points where electron and hole dispersions cross each other, or where the valence and conduction bands become degenerate. The combined 4-fold spin/pseudospin degeneracy gives rise to an emergent SU(4) symmetry, which is useful in analyzing the low-energy properties of graphene and plays a central role in our discussion below. Obviously this is a very unusual band structure, which has attracted much attention both theoretically and experimentally.

The observation of quantum Hall effect (QHE)\cite{novoselov,zhang} when an external, perpendicular magnetic field ($B$) is applied, provides the most compelling evidence for the 2D massless Dirac nature of electrons in graphene. In particular, such a system is predicted\cite{zheng,gusynin,peres} to support integer QHE with quantized values of Hall conductance given by:
\begin{equation}
\sigma_{xy}=\pm g_s g_v (n+1/2) e^2/h,
\label{hall}
\end{equation}
with $n=0, 1, 2, \cdots$ is an integer, and $g_s=g_v=2$ are respectively the spin and pseudospin/valley degeneracies; the latter is inherent in the chiral, massless Dirac equation describing 2D graphene. The half integer form in Eq. (\ref{hall}) arises from the Berry phase associated with the pseudospin index, and the experimental observation of the sequence predicted in the form of Eq. (\ref{hall}) is a direct reflection of the massless chiral Dirac nature of the low-energy electronic states in graphene. Aside from the shift $1/2$, Eq. (\ref{hall}) implies two additional peculiar characteristics of graphene QHE: (i) The $\pm$ sign in front of the quantized Hall conductance indicates the smooth transition from hole-like to electron-like carriers as Fermi energy moves through the Dirac point, as expected from the gapless Dirac spectrum; (ii) the degeneracy factor $g=g_s g_v=4$ is associated with the SU(4) symmetry discussed earlier; in the specific case of massless Dirac electrons moving in an (orbital) magnetic field, it is equivalent to the statement that each Landau level (LL) has a $g=g_s g_v=4$ fold degeneracy, on top of the usual orbital degeneracy.

The graphene LL energy:
\begin{equation}
E_n=\pm\sqrt{2n\hbar v^2|eB|/c}\propto \sqrt{nB},
\label{diracll}
\end{equation}
with $n=0, 1, 2, \cdots$, is yet another key difference between Dirac spectrum as compared to ordinary electron LL spectrum: $E_n=(n+1/2)\hbar\omega_c$ with $\omega_c=eB/(mc)$, where $m$ is {\em effective} mass of the electron (or hole) in a given band. The spectrum of Eq. (\ref{diracll}) has been observed directly in STM measurements.\cite{andrei,miller} We note that the strict 2D nature of graphene (i.e., absence of any higher subbands due to the finite width of quantum wells in usual semiconductor materials) and the large inter LL energy separation for large $B$ and small $n$ has made it possible to observe a room temperature quantized Hall resistance plateau in graphene; in particular, the $\sigma_{xy}=2e^2/h$ plateau is observed with a quantization accuracy of $0.2\%$ at $T=300K$ and $B=45T$.\cite{boebinger}

To facilitate comparison with QHE in regular semiconductor systems, we rewrite Eq. (\ref{hall}) as
\begin{equation}
\sigma_{xy}= \nu e^2/h,
\label{newhall}
\end{equation}
with $\nu=\pm g(n+1/2)$ and $g=g_s g_v=4$. For SU(4) symmetric ground states, we expect QHE at
\beq
\nu=\pm 4n+2= \pm 2, \pm 6, \pm 10, \cdots,
\label{seq}
\eeq
which are indeed the observed QHE sequence at modest magnetic fields; in fact, this sequence has been observed up to $\nu=26$ at $B=9T$ and $T=1.6 K$; this observation decisively reflects the chiral massless Dirac dispersion of graphene with a pseudospin Berry phase.

The question we are interested in pursuing is whether the SU(4) symmetry can be broken in the ground state, leading to the splitting of the 4-fold degeneracy. This lifting has, in fact, been observed at higher magnetic field $B\gtrsim 20 T$, where new QHE has been seen for  $\nu=0,
\pm 1$, and $ \pm 4$.\cite{zhang2} Theoretically this issue has been discussed at great length in the literature.\cite{nomura,yangreview} Various sources of the splittings, including Coulomb interaction, electron spin Zeeman splitting, lattice effects not captured by Dirac equation, and  electron-phonon coupling, have been identified, but no consensus has been reached thus far. Instead of adding to the discussion of the microscopic origin of the splitting, we take a phenomenological approach in this paper, and discuss the consequences of the splitting when present.

To understand these anomalous quantum Hall (QH) states, we introduce spin and valley splittings to the Dirac LL spectrum:
\begin{equation}
E_{n,s,v}=\pm\sqrt{2n\hbar v^2|eB|/c} \pm \Delta_s^n(B) \pm \Delta_v^n(B),
\label{splitting}
\end{equation}
where $\Delta_s^n, \Delta_v^n$ are the spin and valley splittings of the $n$th orbital LL respectively, which depend on both the LL index and magnetic field $B$. The spectrum in the presence of such splittings is illustrated in Fig. 1. For definitiveness we assume $\Delta_v < \Delta_s \ll |E_{n+1} -E_n|$, which are reasonable assumptions for the LLs of interest in graphene.

\begin{centering}
\begin{figure}
\epsfig {file=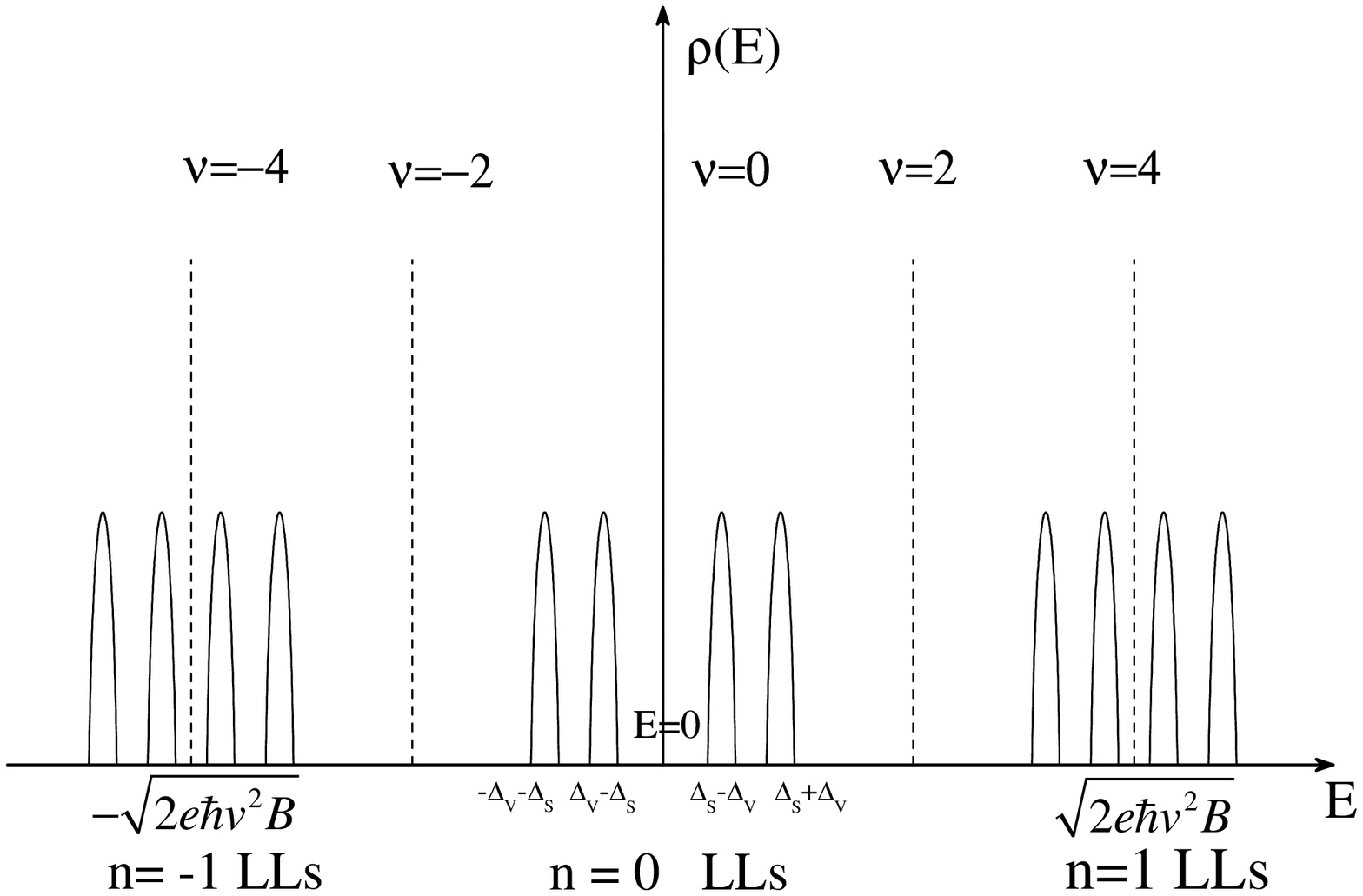,width=80mm}
\epsfig {file=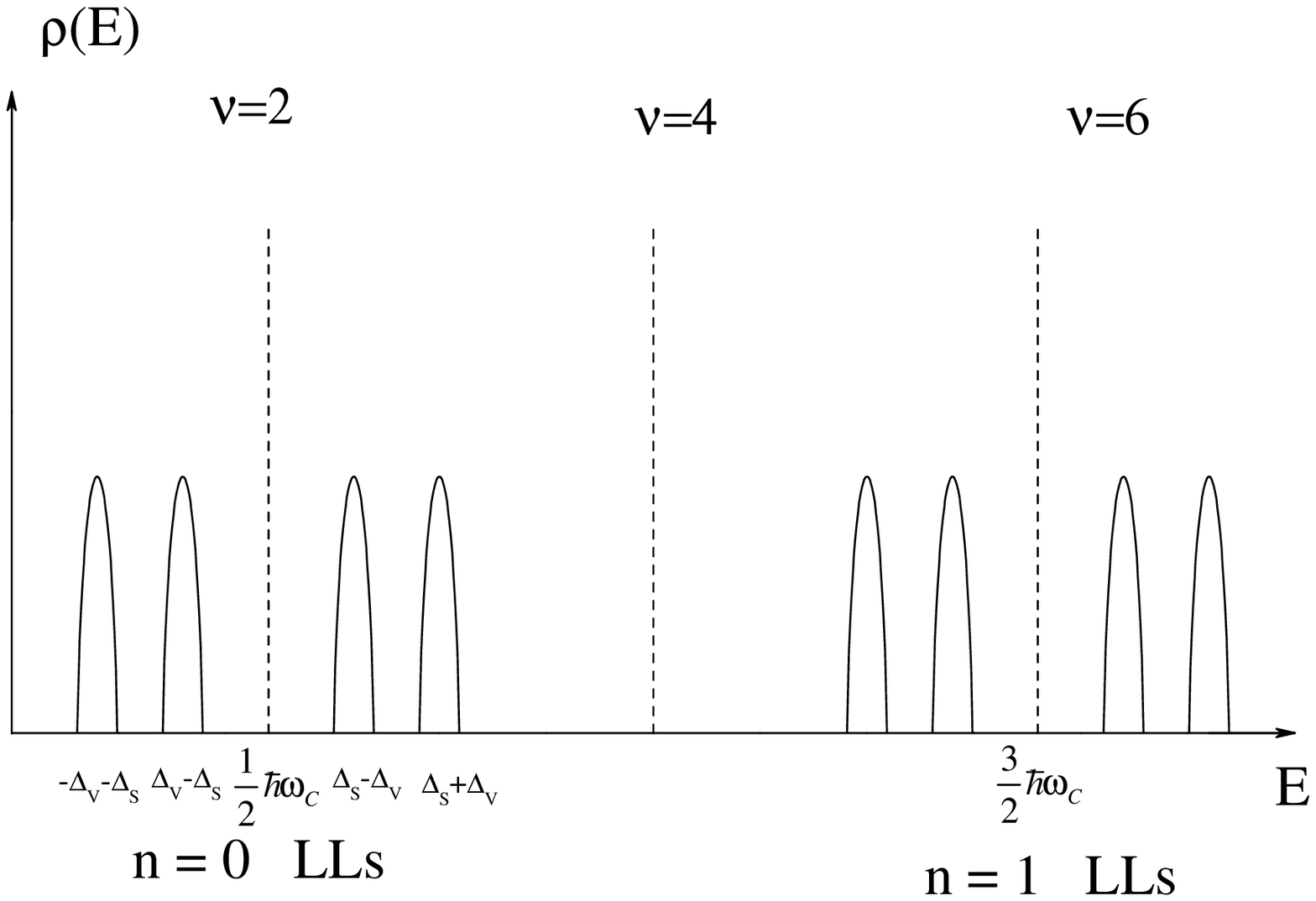,width=80mm}
\caption{Upper panel: Schematic single electron density of states in graphene when the SU(4) symmetry is completely broken. $\Delta_s$ and $\Delta_v$ are the spin and valley splittings respectively. The spin and valley resolved Landau levels (LLs) are further broadened by disorder, with states extended at the center of each LL and localized away from the centers. The dashed lines indicate the locations of the Fermi energies at the center of representative quantum Hall plateaus. Lower panel: Schematic single electron density of states in Si (100) inversion layer when the SU(4) symmetry is completely broken.
}
\label{fig1}
\end{figure}
\end{centering}

We first ask, assuming finite $\Delta_s$ and $\Delta_v$ on phenomenological grounds, what the expected QHE sequence would be for the LL stricture of Eq. (\ref{splitting}) and Fig. 1. Making the standard QHE assumption of the LL tails being occupied by localized states and the LL centers being extended, we conclude that, if all the 4-fold degenerate LLs are indeed split into distinct spin and valley split individual levels, we will have the sequence of QHE with
\beq
\nu=0, \pm 1, \pm 2, \cdots,
\label{splitseq}
\eeq
instead of the sequence of $\nu=\pm 4n+2= \pm 2, \pm 6, \pm 10, \cdots$ with SU(4) symmetry present in QH ground states. If we assume $\Delta_s > 0$ while $\Delta_v=0$, so the symmetry reduces from SU(4) to SU(2), the corresponding sequence would be
\beq
\nu=0, \pm 2, \pm 4, \cdots.
\eeq
The same would be true in the opposite case with $\Delta_s = 0$ while $\Delta_v > 0$.

The quantization of $\sigma_{xy}$ is associated with the vanishing of longitudinal conductivity $\sigma_{xx}$ in QHE, as can easily be seen from Laughlin's gauge arguments (and augmented by Halperin).\cite{laughlin} Thus corresponding to
any of the QH sequence discussed here, we also have
\beq
\sigma_{xx}=0
\eeq
on the QH plateaus, at $T=0$. We emphasize that the quantization of $\sigma_{xy}$ and vanishing of $\sigma_{xx}$ in the limit $T\rightarrow 0$ are intrinsically tied through gauge invariance, as is most clearly demonstrated by the Laughlin gauge arguments; these properties in fact {\em define} QHE.

The main difference between 2D graphene and ordinary 2D semiconductor systems with parabolic or non-relativistic electron dispersion (e.g., Si MOSFETs, GaAs heterostructures and quantum wells) is {\em not} the 4-fold spin-pseudospin degeneracy, but the presence of the $n=0$ LL at $E=0$ which has equal weight in the electron and hole bands. It is instructive to compare the LL structure of graphene with that of the Si(100) inversion layer,\cite{stern} where the LL energies are
\beq
E_n=(n+1/2)(\hbar eB/mc),
\eeq
with $n=0, 1, 2, \cdots$, while the QH sequence is given by
\beq
\nu=gn=4, 8, 12, \cdots.
\eeq
We note that Si(100) MOSFET electrons also have an SU(4) spin-valley symmetry, and consequently, each LL is $g$-fold degenerate with $g=g_sg_v=4$; yet the SU(4) symmetric QH sequence above is {\em different} from that of graphene, Eq. (\ref{seq}). Here the spin degeneracy is the same as in graphene, but the valley degeneracy, $g_v=2$, arises from the bulk Si band structure which has six equivalent ellipsoid minima (i.e., ``valleys") in the conduction band close to the edges of the Brillouin zone; this is in contrast to the graphene case, where the valley degeneracy is an inescapable consequence of its Dirac nature. In a strong magnetic field the spin and valley degeneracies of Si(100) MOSFET are also lifted, giving rise to the LL spectrum (c.f. Eq.(\ref{splitting}) for graphene, and see Fig. 1):
\beq
E_n=(n+1/2)(\hbar eB/mc)\pm \Delta_s^n(B) \pm \Delta_v^n(B).
\eeq
This would lead to the familiar integer QH sequence $\nu=1, 2, 3, \cdots$. Comparing with that of graphene Eq. (\ref{splitseq}) when SU(4) symmetry is completely broken, other than the negative $\nu$ states corresponding to hole QH states, the biggest difference is the presence of $\nu=0$ QH state in graphene, which is the focus of this paper.

It is instructive to discuss further the phenomenology of QHE in Si inversion layers in the context of understanding graphene QHE since both systems in their pristine states have spin and pseudospin (i.e. valley) symmetries. The most extensively studied Si QHE is in Si (100) inversion layers where the original QHE discovery was made by von Klitzing.\cite{klitzing}  This 2D system, as mentioned above, has an SU(4) symmetry, which is found to be lifted in the low-lying LLs leading to $\nu=1,2,3,4,\cdots$ quantization in the observed QHE $\rho_{xy}=h/(\nu e^2)$.  We emphasize that in Si (100) 2D system, the valley splitting is much less than the spin splitting, and in fact, in higher orbital LLs the valley splitting is typically not manifested in the observed QHE which follow the SU(2) sequence with $\nu=2, 4, 6, 8, \cdots$ for larger orbital levels.  This is again similar to graphene where valley splitting effects have so far seemed to have been observed only in the lowest LL.  Similar to graphene, the origin of valley splitting in Si (100) system is still not theoretically well-understood,\cite{Saraiva} and there is not much direct experimental evidence for valley splitting in the absence of a magnetic field. It is conceivable that in graphene the valley splitting is strongly enhanced by interaction effects in a magnetic field leading to the full lifting of the SU(4) symmetry.  The phenomenological parallel between graphene and Si QHE with respect to the ground state SU(4) symmetry breaking points toward the possible importance of interaction effects playing a dominant role in lifting the valley degeneracies in both cases.  Finally, we point out that in Si (111) inversion layers, the ground state valley degeneracy is six, leading to an effective SU(12) ground sate symmetry. Random uniaxial interfacial stress can lift this SU(12) symmetry converting it to an SU(4) ground state symmetry by splitting the six-fold valley degeneracy into two lower valleys and four upper valleys.  Therefore, Si (111) system becomes equivalent to the Si (100) SU(4) situation, with the possible further lifting of the SU(4) symmetry in high fields at the lowest LLs. QHE has been seen in Si (111) experimentally.\cite{kane}


In Si MOSFET and other non-relativistic 2D electron systems, for sufficiently high $B$ (where all electrons are in the $n=0$ LL with filling factor $\nu\ll 1$), the system enters an insulating regime with longitudinal resistivity $\rho_{xx}\rightarrow\infty$ as $T\rightarrow 0$, while the Hall resistivity $\rho_{xy}$ taking its classical (and roughly temperature-independent) value. It is worth noting that experimentally, one normally uses the Hall bar geometry to directly measure $\rho_{xx}$ and $\rho_{xy}$ (actually to be more precise, longitudinal and Hall resistances $R_{xx}$ and $R_{xy}$, and convert them to $\rho_{xx}$ and $\rho_{xy}$ with appropriate geometrical factors); $\sigma_{xx}$ and $\sigma_{xy}$ can then in turn be obtained through the tensor inversion relations:
\beqarr
\sigma_{xx}=\rho_{xx}/(\rho_{xx}^2+\rho_{xy}^2);\hskip 0.3cm \sigma_{xy}=\rho_{xy}/(\rho_{xx}^2+\rho_{xy}^2).
\label{sigma}
\eeqarr
It is clear that in this high $B$ insulating phase (often called the Hall insulator),\cite{hwjiang} one has $\sigma_{xy}\rightarrow 0$ {\em and} $\sigma_{xx}\rightarrow 0$ in the limit $T\rightarrow 0$, for a non-zero range of $B$; this, by definition, corresponds to a $\nu=0$ QH state! But in that context this high $B$ insulating phase is {\em not} referred to as a QH state, mainly because {\em no} plateau is seen in $\rho_{xy}$, which is what is measured directly, despite the fact that there {\em is} a plateau in $\sigma_{xy}$, as we argued above. This relates to our discussion about the important difference between QH states with $\nu=0$ and $\nu\ne 0$.

The tensor inversions of Eqs. (\ref{sigma}) are
\beqarr
\rho_{xx}=\sigma_{xx}/(\sigma_{xx}^2+\sigma_{xy}^2);\hskip 0.3cm \rho_{xy}=\sigma_{xy}/(\sigma_{xx}^2+\sigma_{xy}^2).
\label{rho}
\eeqarr
As a result for any QH state with $\nu\ne 0$, we have
\beqarr
\rho_{xx}=0;\hskip 0.3cm \rho_{xy}=1/\sigma_{xy}=h/(\nu e^2).
\label{rho1}
\eeqarr
These equations are deceptively simple, and form the basis of studying QHE -- the deceptive dichotomy here, often lost in the literature, is that theoretically the fundamental quantities are $\sigma_{xx}(=0)$ and $\sigma_{xy}(=\nu e^2/h)$, whereas experimentally the fundamental (or more specifically, the measured) quantities are $\rho_{xx}$ and $\rho_{xy}$.

The $\sigma, \rho$ dichotomy has not been an issue in QHE studied in semiconductor based 2D systems because there one always has
\beq
\sigma_{xy}\gg \sigma_{xx}\sim 0
\eeq
in the quantization regime, leading to Eqs. ({\ref{rho1}); thus the longitudinal transport coefficients vanish and Hall coefficients quantized, in {\em both} $\sigma$ and $\rho$.

In graphene, however, the $n=0$ LL resides precisely at the particle-hole symmetric point of $E=0$, which allows for a $\sigma_{xy}=0$ QHE when the SU(4) symmetry is broken, as indeed observed experimentally.\cite{zhang2,jiang,ong1,ong2,giesbers,lzhang} Putting $\sigma_{xy}=0$ in Eqs. (\ref{rho}) we get
\beqarr
\rho_{xy}=0;\hskip 0.3cm \rho_{xx}=1/\sigma_{xx}.
\label{rho2}
\eeqarr
But, the quantization of Hall conductivity, even when the quantized value is zero, demands $\sigma_{xx}\rightarrow 0$ as $T\rightarrow 0$, and consequently we have for the $\nu=0$ graphene QHE
\beqarr
\rho_{xx}=1/\sigma_{xx} \rightarrow\infty.
\label{rho3}
\eeqarr

As discussed earlier, this is very similar to what happens in the high field insulating phase (or Hall insulator phase) in ordinary 2D electron gas with parabolic band structure. The parallel between them is actually quite complete.  In particular, the experimental Hall insulator phase has $\rho_{xx}$ going to infinity and $\rho_{xy}$ given by the classical formula, $\rho_{xy}=B/Nec$, where $N$ is the 2D carrier density in the system.  As we discussed above, an inversion of the $\rho$ tensor leads to the Hall insulator conductivity $\sigma_{xx}=0$ and $\sigma_{xy}=0$, which is formally the same as the $\nu=0$ QH conductivity values around the Dirac point.  Now we note that the corresponding graphene Hall insulator phase must necessarily have $\rho_{xy}=0$ since the 2D carrier density at the Dirac point (i.e. at the zero energy charge neutrality point), separating the electron and hole bands, is by definition zero (i.e. $N=0$ in the classical Hall resistivity formula). This shows that the emergent divergent longitudinal resistivity at the graphene Dirac point could equally well be considered a putative Hall insulator phase, which in this context, is equivalent to the zero Fermi energy $\nu=0$ QHE in graphene with $\sigma_{xx}=0$ and $\rho_{xy}\sim 0$.  The only (qualitative) difference between graphene and 2D semiconductor systems is that the graphene Dirac point is known to be dominated by density inhomogeneities associated with the electron-hole puddles,\cite{adam,hwang07,martin,rossi08,Deshpande}
which lead to considerable density fluctuations around the average (expected) zero density at the Dirac point.  This means that there will be considerable fluctuations around the expected $\rho_{xy}=0$ value in the graphene Hall insulator phase. If these fluctuations are large, then neither the divergent resistivity nor the $\nu=0$ graphene QH (or the equivalent Hall insulator phase) would be observable, indicating the need for very high mobility samples where these fluctuations are suppressed\cite{rossi08} as well as very high magnetic fields so that the magnetic length is well below the typical puddle  size.  We emphasize that in this bulk picture that we are proposing there is no finite-$T$ phase transition in the system, only a possible $T=0$ quantum phase transition to the $\nu=0$ phase.  At finite $T$ and in the presence of strong edge effects (i.e. mesoscopic samples), there may be substantial modification of this underlying picture, but we think that it is unlikely that there will be any universal physics in such a situation and details (such as the mechanism producing the SU(4) symmetry breaking) will play crucial roles.

Based on very general Laughlin-type gauge considerations applied to bulk 2D graphene QHE, we thus arrive at the result summarized below: If the SU(4) spin/valley symmetry of graphene ground state is lifted, then there is a $\nu=0$ QH state with $\sigma_{xy}=0$ when the Fermi energy is at (and around) the charge-neutral Dirac point, with a divergent {\em bulk} longitudinal resistivity $\rho_{xx}\rightarrow\infty$ when $T\rightarrow 0$; this is a necessary consequence of having a $\nu=0$ QH plateau. We thus believe the recent observation\cite{ong1,ong2,giesbers,lzhang} of divergent resistivity at the graphene neutral point in a high magnetic field is a direct manifestation of the SU(4) symmetry-broken $\nu=0$ QHE.

\begin{centering}
\begin{figure}
\epsfig {file=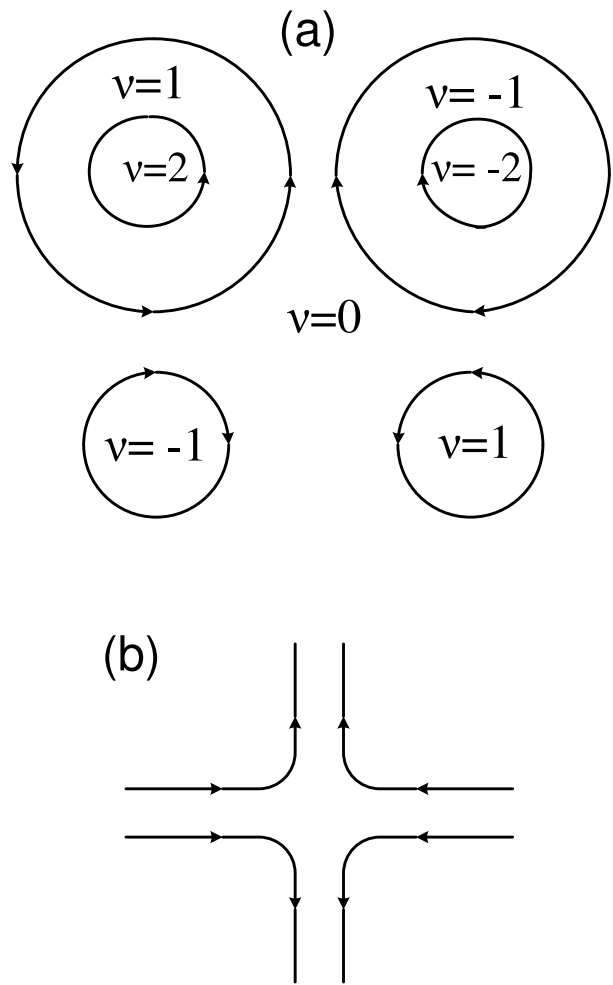,width=80mm}
\caption{(a) Schematic illustration of formation of domains with different $\nu$'s due to electron and hole puddles, and current carrying states flowing along the domain walls. (b) A building block (saddle point) of an appropriate network model describing these domain wall states.
}
\label{fig2}
\end{figure}
\end{centering}

In the following we make a few more comments about this conclusion, in which some specific issues related to experimental studies will be addressed.

(i) All our arguments presented above are about the zero-temperature bulk physics in the thermodynamic limit; technically, this implies that the thermodynamic limit is taken {\em before} any other limits, in particular, the zero temperature limit $T\rightarrow 0$. As a result mesoscopic effects like edge transport play no role in our considerations. It is conceivable however, that small graphene samples used in experiments may be in the mesoscopic regime at sufficiently low temperatures, and edges may play an important or even a dominant role in transport; transport theories based on edge channels have been developed.\cite{abanin,fertig} It is also conceivable that differences\cite{zhang2,jiang,ong1,ong2,giesbers,lzhang,abanin} observed in the transport properties near the Dirac point may also be due to the differences in edge properties of different samples. This leads us to a straightforward prediction: If one measures the {\em bulk} conductivity directly using the Corbino geometry, one should get $\sigma_{xx}\rightarrow 0$ in {\em all} samples exhibiting the $\nu=0$ QHE, due to the absence of edge contribution to transport.

(ii) Our bulk arguments are completely independent of the details of how the SU(4) symmetry is broken. It could be due to a spontaneous symmetry breaking driven by Coulomb interaction, or an explicit symmetry breaking induced by Zeeman splitting or intervalley scattering induced valley splitting. On the other hand the mesoscopic effects discussed above {\em are} sensitive to the specific manner in which the symmetry is broken, see, {\em e.g.}, Ref. \onlinecite{abanin} and on whether the valley-splitting or the spin-splitting is larger in magnitude.

(iii) Our bulk consideration suggests that $\rho_{xx}\rightarrow\infty$ does {\em not} reflect a {\em bulk} phase transition driven by $B$ at {\em finite} $T$, as it is a property of the underlying QH phase. There {\em can} be a quantum phase transition at $T=0$, driven by $B$, into the $\nu=0$ QH phase.\cite{note} Temperature can, however, be very important in the role played by the above mentioned mesoscopic effects.

(iv) It is known,\cite{adam,martin} as discussed above, that due to the presence of disorder arising from charged impurities and the lack of screening due to vanishing density of states at the Dirac point, neutral graphene samples are quite inhomogeneous, and electron and hole puddles form at $B=0$. Due to such density inhomogeneity, it is quite possible that the observation of the $\nu=0$ QHE in graphene necessitates not only the complete lifting of the SU(4) symmetry in the $n=0$ graphene LL, but also a sufficiently high magnetic field. This is because in the presence of $B$, these electron and hole puddles will turn into domains of different (integer) $\nu$'s, with current carrying states propagating along the domain walls; see Fig. 2a. Such domains and domain walls on top of the $\nu=0$ background (or percolating domain), when present (and especially if large), can dominate {\em bulk} transport, thus suppressing the divergent $\rho_{xx}$. The situation is very different for $\nu\ne 0$, where the background or percolating domain has a non-zero Hall conductance, thus dominating bulk transport. As $B$ increases, we expect the symmetry breaking splittings $\Delta_s$ and/or $\Delta_v$ (that give rise to the $\nu=0$ QHE) to increase correspondingly. This will lead to suppression and eventual elimination of these domains. We thus conjecture the critical field observed in Refs. \onlinecite{ong1,ong2} beyond which very strong insulating behavior kicks in corresponds to the field that leads to such suppression. This is clearly consistent with the observation that the more disordered the sample (as measured by mobility), the higher the critical field , as the electron hole puddles or domains are due to disorder. Theoretically, one can formulate a network model to describe transport through these domain wall states; see Fig 2b. It turns out the appropriate network model takes the same form as that of a system with random but zero average magnetic flux.\cite{kim} This system was argued to support a Kosterliz-Thouless (KT) type metal-insulator transition.\cite{arovas} A number of numerical works indeed find evidence of such a transition,\cite{liu} although it was argued later on\cite{aronov,kim} that actually all states are localized, albeit with extremely large localization length at weak disorder, giving rise to an apparent transition in finite size numerical studies. Such an apparent transition may be related to the apparent KT-like transition seen in Refs. \onlinecite{ong1,ong2}.  We therefore suggest that scanning probe measurements \cite{martin,Deshpande} be carried out in graphene samples under QHE conditions to directly probe the electron-hole puddles and the associated domains which may be dominating the experimental QH physics around the Dirac point.

We conclude by asserting that theoretical considerations of the $\nu=0$ QHE in graphene using Laughlin's gauge argument lead to a divergent bulk $\rho_{xx}$ associated with the quantization of $\sigma_{xy}$, when the Fermi energy is at the Dirac point. This is an intrinsic bulk property of the system, which can be viewed both as a quantized Hall liquid and a Hall insulator. In our consideration edges play no role, and there does not need to be a bulk phase transition associated with the the divergence of $\rho_{xx}$. A clear and simple prediction is that {\em all} systems that exhibit a quantization of $\sigma_{xy}=0$ will have $\sigma_{xx}\rightarrow 0$ in a Corbino disk geometry measurement. We also suggest that the experimentally observed magnetic field-driven insulating behavior is due to the suppression of domains or electron and hole puddles by field.  These domains should be directly observable in scanning probe measurements.

We thank Wenxin Ding for assistance with the figures. This work is supported by Microsoft Station Q (SDS), DARPA-QuEST (SDS), NSF-NRI (SDS), and NSF-DMR-0704133 (KY). Part of this work was carried out while the authors were visiting Kavli Institute for Theoretical Physics (KITP). The work at KITP is supported in part by National Science Foundation grant No. PHY-0551164.

\end{document}